\documentstyle[12pt,appbwb,a4]{article}

\input psfig

\preprint{\rm Presented at the workshop on
{\em Few-quark problems}, Bled (Slovenia), 8-15 July 2000}

\begin{document}

\title{{\large {\bf Distinct Hagedorn temperatures from particle spectra:\\
a higher one for mesons, a lower one for baryons}}\thanks{%
Research supported in part by the Scientific and Techological Cooperation
Joint Project between Poland and Slovenia, financed by the Ministry of
Science of Slovenia and the Polish State Commettee for Scientific Research,
and by the Polish State Committee for Scientific Research, project
2~P03B~094~19}}
\author{Wojciech BRONIOWSKI\thanks{%
E-mail: broniows@solaris.ifj.edu.pl} 
\address{The H. Niewodnicza\'{n}ski Institute of Nuclear Physics,
         PL-31342 Cracow, POLAND}}
\date{}
\maketitle

\begin{abstract}
We analyze experimental particle spectra and show that the Hagedorn
temperature is significantly larger for mesons than for baryons. The effect
can be explained within dual string models: excitations of three strings in
the baryon produce ``faster'' combinatorics than a single string in the
meson, hence lead to a more rapid growth of baryons than mesons. Predictions
of other approaches for the gross features of particle spectra are also
discussed.
\end{abstract}

\bigskip \bigskip \bigskip \bigskip

{\em This research is being carried out in collaboration with Wojciech
Florkowski and Piotr \.{Z}enczykowski from INP, Cracow.}

\bigskip\bigskip \bigskip

\section{Introduction}

The famous Hagedorn hypothesis \cite{hagedorn,SBM1,hag94}, dating back to
pre-chromodynamic times of the sixties, states that at asymptotically large
masses, $m$, the density of hadronic resonance states, $\rho (m)$, grows
exponentially: 
\begin{equation}
\rho (m)\sim \exp \left( \frac{m}{T_{H}}\right)  \label{hag}
\end{equation}
The Hagedorn temperature, $T_{H}$, is a scale controlling the exponential
growth of the spectrum. Although the Hagedorn hypothesis has sound
thermodynamical consequences (one cannot heat-up a hadronic system above
this temperature), $T_{H}$ should not be immediately associated with
thermodynamics. In this talk we are concerned with the spectrum of particles 
{\em per se}, as read off form the Particle Data Tables \cite{PDG}. In this
context the ``temperature'' $T_H$ is just a parameter in Eq.~(\ref{hag}).

Ever since hypothesis (\ref{hag}) was posed, it has been believed that there
is one universal Hagedorn temperature for all hadrons. {\em Presently
available experimental data show that this is not the case}, as has been
pointed out by W. Florkowski and WB in Refs. \cite{twoT,tmb}.

\begin{figure}[tbp]
\centerline{%
\psfig{figure=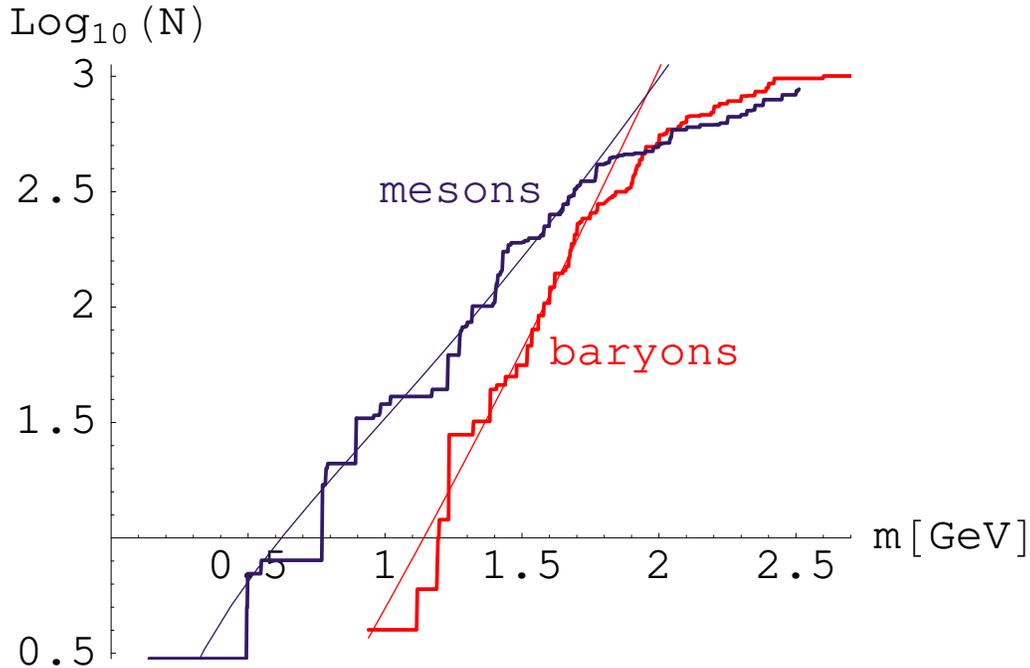,height=9cm,bbllx=82bp,bblly=384bp,bburx=531bp,bbury=680bp,clip=}}
\vspace{0mm} \label{cumul}
\caption{Cumulants of meson and baryon spectra, and the Hagedorn-like fit
with Eqs.~(\protect{\ref{eee}},\protect{\ref{HR}}), 
plotted as functions of mass.}
\end{figure}

This talk has two parts: experimental and theoretical. In the experimental
part (Sec. \ref{experiment}) we show how well the Hagedorn hypothesis works
even for very low masses, and point out the key observation that {\em the
mesonic temperature is significantly larger from the baryonic temperature}.
In the theoretical part (Sec. \ref{theory}) we argue that the only framework
(known to us) which is capable of producing the observed behavior in a
natural way are the Dual String Models \cite{Jacob}. In Sec. \ref{otherapp}
we discuss other approaches and more speculative ideas.

\section{Experiment}

\label{experiment}

\subsection{Experimental spectra of mesons and baryons}

\begin{figure}[tbp]
\centerline{%
\psfig{figure=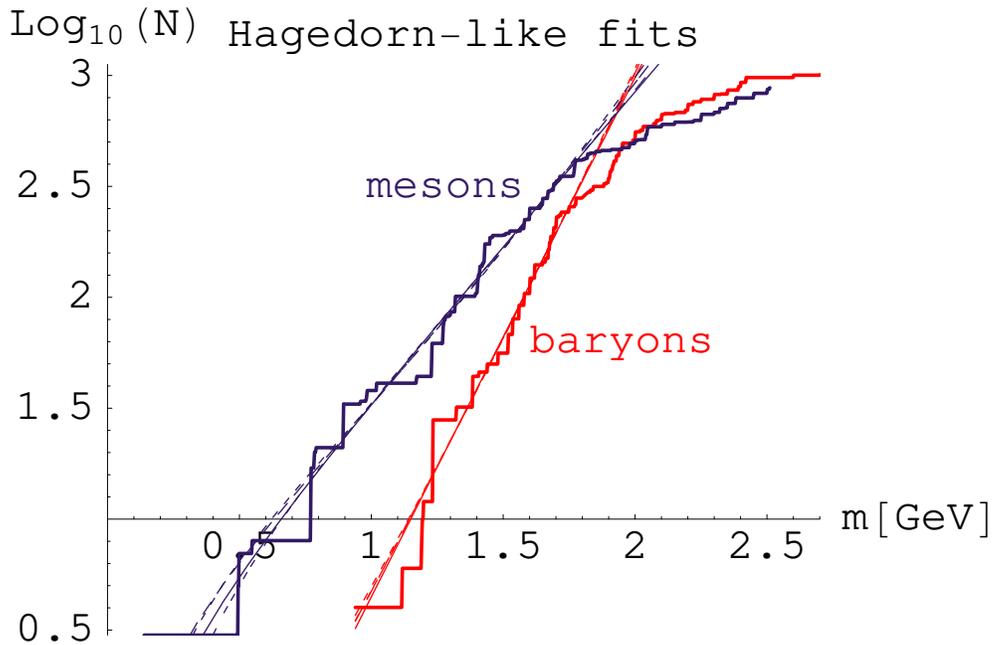,height=9cm,bbllx=23bp,bblly=450bp,bburx=480bp,bbury=758bp,clip=}}
\vspace{0mm}
\caption{Various Hagedorn-like fits, made according to formulas of Table 
\protect{\ref{tabhag}}.}
\label{varhag}
\end{figure}

In Fig. \ref{cumul} we compare the {\em cumulants} of the spectrum \cite{PDG}%
, defined as the number of states with mass lower than $m$. The experimental
curve is 
\begin{equation}
N_{{\rm exp}}(m)=\sum_{i}g_{i}\Theta (m-m_{i}),
\end{equation}
where $g_{i}=(2J_{i}+1)(2I_{i}+1)$ is the spin-isospin degeneracy of the $i$%
th state, and $m_{i}$ is its mass. The theoretical curve corresponds to 
\begin{equation}
N_{{\rm theor}}(m)=\int_{0}^{m}\rho _{{\rm theor}}(m^{\prime })dm^{\prime },
\end{equation}
where 
\begin{equation}
\rho _{{\rm theor}}(m)=f(m)\exp (m/T),  \label{eee}
\end{equation}
with $f(m)$ denoting a slowly-varying function. A typical choice \cite
{hag94,HagRan}, used in the plot of Fig. \ref{cumul}, is 
\begin{equation}
f(m)=A/(m^{2}+(500{\rm MeV})^{2})^{5/4}.  \label{HR}
\end{equation}
Parameters $T_{H}$ and $A$ are obtained with the least-square fit to $\log
N_{{\rm theor}}$, made over the range up to $m=1.8{\rm GeV}$, and skipping
the lightest particle in the set. Other choices of $f(m)$ give fits of
similar quality (see Fig.~\ref{varhag}). A striking feature of Fig. \ref
{cumul} is the linearity of $\log N$ starting at very low $m$, and extending
till $m\sim 1.8{\rm GeV}$. Clearly, this shows that (\ref{hag}) is valid in
the range of available data.\footnote{%
Above 1.8GeV the data seems to be sparse and we should wait for this region
to be explored by future experiments.} However, the slopes in Fig. \ref
{cumul} are {\em different} for mesons and baryons. For the assumed $f(m)$
of Eq. (\ref{HR}) we get 
\begin{equation}
T_{{\rm meson}}=195{\rm MeV},\;\;\;T_{{\rm baryon}}=141{\rm MeV}.  \label{Ts}
\end{equation}
This means that $T_{{\rm meson}}>T_{{\rm baryon}}$, and the inequality is
substantial! Although it has been known to researchers in the field of
hadron spectroscopy that the baryons multiply more rapidly than mesons \cite
{freund}, to our knowledge this fact has not been presented as vividly as in
Fig. \ref{cumul}. To emphasize the strength of the effect we note that in
order to make the meson line parallel to the baryon line, we would have to
aggregate $\sim 500$ additional meson states up to $m=1.8$MeV as compared to
the present number of $\sim 400$.

\subsection{Are we asymptotic?}

An important question is whether the presently available range of masses is
asymptotic in view of Eq. (\ref{hag}). The answer is {\em no}! This is how
we can look at this question quantitatively.
\begin{table}[tb]
\begin{centering}
\begin{tabular}{|c|c|c|c|c|c|}
\hline
Formula & $m_{0}$  &  $T_{\rm{mes}}$ & $T_{\rm{bar}}$ &
 $\sigma^2_{\rm{mes}}$ & $\sigma^2_{\rm{bar}}$ \\
& MeV & MeV & MeV & & \\
\hline
\hline
{$\frac{A}{(m^2+m_0^2)^{5/4}}\exp ({m\over T})$} & 500
& 195 & 141 & 0.016 & 0.015 \\
\hline
- - - & 1000 & 228 & 152 & 0.014 & 0.015  \\
\hline
- - - & 250  & 177 & 136 & 0.025 & 0.015 \\
\hline
{$\frac{A}{(m+m_{0})^{5/2}}\exp (\frac{m}{T})$} &
 1000 & 223 &
154 & 0.015 & 0.015 \\
\hline
{$A \exp (\frac{m}{T})$} &  & 311 & 186 & 0.014 & 0.015 \\
\hline
{$\frac{A}{m}I_{2}(\frac{m}{T})$} &
 & 249 & 157 & 0.014 & 0.015\\
\hline
\end{tabular}
\medskip
\caption{Various Hagedorn-like fits. Rows 1-4 use formulas of
Ref. \protect{cite{SBM1}}, row 5 uses a simple exponent, and row
6 uses the scalar string model of Ref. \protect{\cite{dienes}}.
The last two column display the mean suared deviation for
the meson and baryon case, respectively.}
\label{tabhag}
\end{centering}
\end{table}
Consider the generic form of the spectrum of Eq. (\ref{eee}). We can rewrite
it as 
\begin{eqnarray}
&&f(m)e^{m/T}=e^{\log f(m)+m/T}\simeq e^{\log [f(\overline{m})+f^{\prime }(%
\overline{m})\Delta m]+(\overline{m}+\Delta m)/T}=  \nonumber \\
&&{\rm const}\;e^{\left( \frac{1}{T}+\frac{f^{\prime }(\overline{m})}{f(%
\overline{m})}\right) \Delta m}={\rm const}\;e^{\frac{\Delta m}{T_{{\rm eff}}%
}},  \nonumber  \label{asi}
\end{eqnarray}
where $m=\overline{m}+\Delta m$, and in the range of data $\overline{m}\sim
1 $GeV. We have defined $T_{{\rm eff}}$ as the {\em effective} Hagedorn
temperature in the (non-asymptotic) region around $\overline{m}$. The value
of $T_{{\rm eff}}$ follows directly from the data. We have, according to Eq.
(\ref{asi}), 
\begin{equation}
\frac{1}{T}=\frac{1}{T_{{\rm eff.}}}-\frac{f^{\prime }(\overline{m})}{f(%
\overline{m})}.
\end{equation}
The following statements are obvious:

\begin{itemize}
\item  since $f^{\prime}(\overline{m})<0$, $T<T_{{\rm eff}}$,

\item  only at $m \to \infty$ we have $T=T_{{\rm eff}}$. In the region of
data we find significant differences between $T$ and $T_{{\rm eff}}$.
\end{itemize}

Here is a numerical example. Consider 
\begin{equation}
f(m)=\frac{A}{(m^2+m_0^2)^{5/4}},
\end{equation}
which leads to 
\begin{equation}
\frac{1}{T}=\frac{1}{T_{{\rm eff.}}}+\frac{5}{2} \frac{\overline{m}}{(%
\overline{m}^2+m_0^2)}
\end{equation}
Now we take $m_0=0.5$GeV and $\overline{m}=1$GeV and find

\begin{itemize}
\item[~]  ~for~mesons: $T_{{\rm eff}}=311$MeV, $T=192$MeV (exact fit: 195MeV)

\item[~]  for~baryons: $T_{{\rm eff}}=186$MeV, $T=136$MeV (exact fit: 141MeV)
\end{itemize}

We conclude that only in the asymptotic region, $m >> m_0$, the choice of $%
f(m)$ is not important. In the region of presently-available data $f(m)$
matters very much for the extracted values of the Hagedorn temperature. This
simply means that we need a {\em theory} in order to make quantitative
statements!

The numerical parameters obtained from various choices of the function $f(m)$
are collected in Table \ref{tabhag}. Figure \ref{varhag} shows the fits
corresponding to the rows 1, 4, 5 and 6 of Table \ref{tabhag}. Note the fits
are very close to each other and the theoretical curves are virtually
indistinguishable in the region of data. In view of the above discussion it
makes little sense to treat the Hagedorn temperature as an absolute
parameter and to quote its value without specifying the model that yields
the function $f(m)$.

\subsection{Flavor universality}

\begin{figure}[tbp]
\centerline{%
\psfig{figure=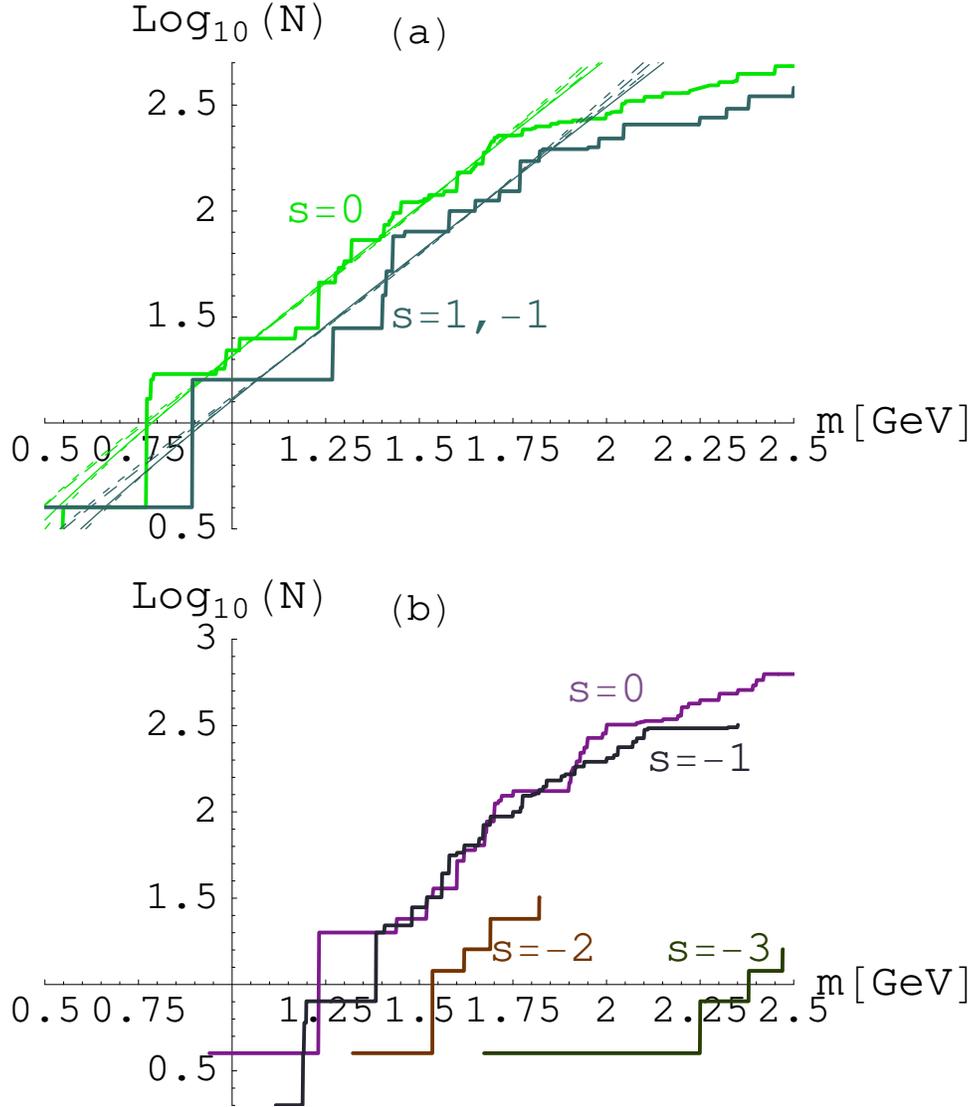,height=15cm,bbllx=30bp,bblly=262bp,bburx=480bp,bbury=758bp,clip=}}
\vspace{0mm}
\caption{Strange vs. non-strange mesons (a), and baryons (b).}
\label{sns}
\end{figure}

In Fig. \ref{sns} we show the cumulants of particle spectra of a given value
of strangeness. We can clearly see that the slopes in the figure do not
depend on strangeness. The meson plot includes various Hagedorn fits of Fig. 
\ref{varhag}. The two sets of lines are displaced in the $m$ variable by
roughly $150$MeV, which is the difference of the masses on the strange and
non-strange quarks. The conclusion here is that the addition of the strange
quark mass has no effect on the rate of growth of the number of states with $%
m$. Certainly, we are rediscovering the $SU(3)$ flavor symmetry here!

\subsection{Plot in the exponential variable}

\begin{figure}[tbp]
\centerline{%
\psfig{figure=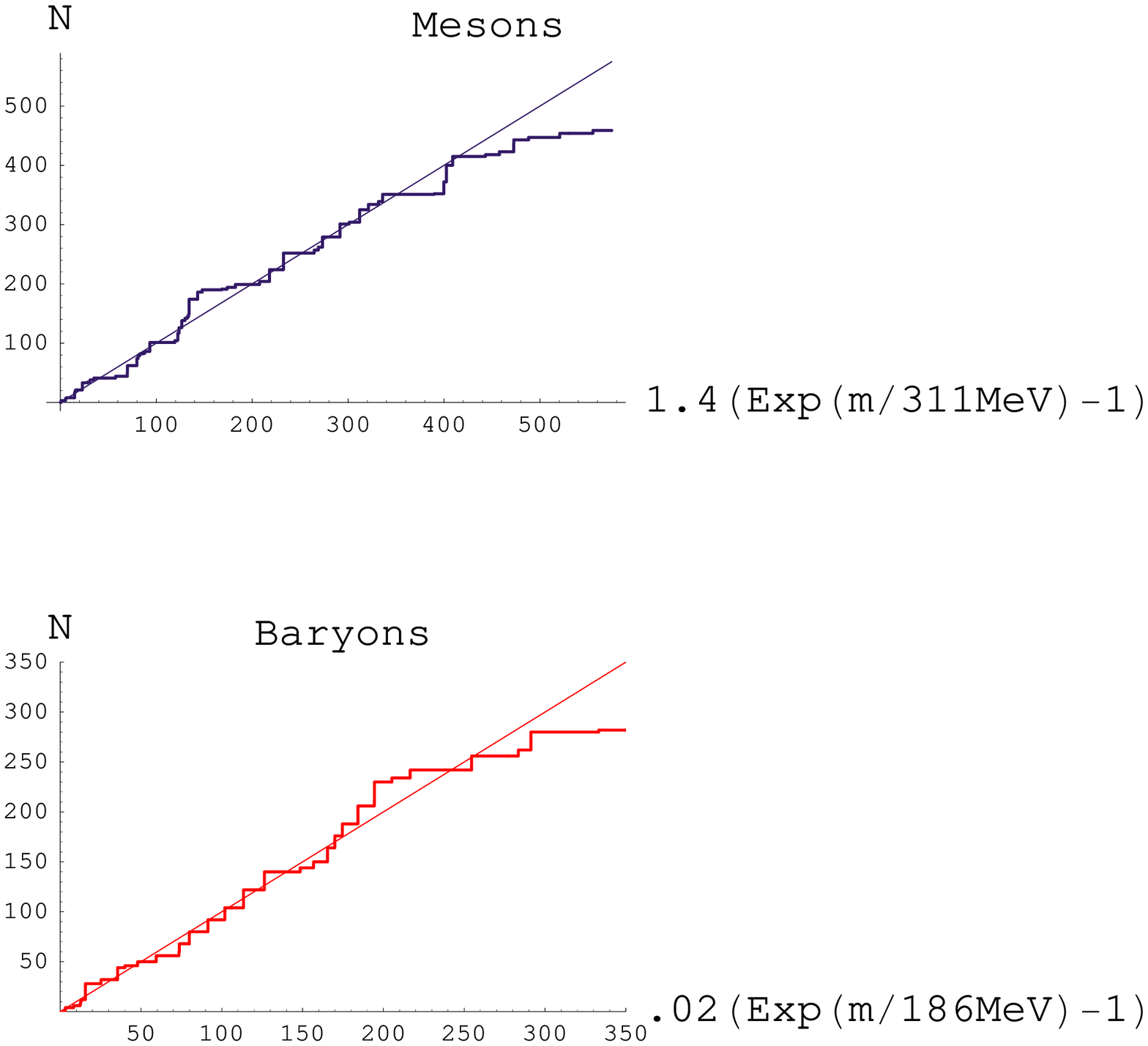,height=13cm,bbllx=34bp,bblly=200bp,bburx=583bp,bbury=713bp,clip=}}
\vspace{0mm}
\caption{Cumulants of the meson and baryon spectra plotted in exponential
variables.}
\label{expo}
\end{figure}

We end the experimental part of this talk by showing the same information as
in Fig. \ref{cumul}, but instead of using logarithmic units on the vertical
axes, we take exponential units on the horizontal axis. More precisely, we
take the fit to the spectrum with of the form with the simple exponent (row
5 in Table \ref{tabhag}), which leads to the cumulant $N(m)=A T
(\exp(m/T)-1) $, where the values of $A$ and $T$ result from the
least-square fit. Next, we define the variable $y=A T (\exp(m/T)-1)$ and
plot the cumulants as functions of $y$. Note that the $A$ and $T$ parameters
are different for mesons and baryons. Again, the linearity of data in the
figure is striking. It starts at basically $m=0$, and extends to $m \sim 1.8$%
GeV. The advantage of the plot in Fig. \ref{expo} to that of Fig. \ref{cumul}
is that now the steps in the experimental cumulant are of a similar size
independently of $m$.

\bigskip

We conclude this section by stating that the exponential growth of hadronic
spectra in the region of $m$ up to about 1.8GeV, with $T_{{\rm mes}} > T_{%
{\rm bar}}$, is an {\em experimental fact}.

\section{Theory}

\label{theory}

We are faced with two basic theoretical questions:

\begin{enumerate}
\item  {\em Why is the spectrum of resonances exponential?}

\item  {\em Why do mesons and baryons behave so differently?}
\end{enumerate}

Concerning the first question, let us stress that it is not at all easy to
get an exponentially rising spectrum of resonances. Take the simplistic
harmonic-oscillator model, whose density of states grows as $m^{d-1}$, with $%
d$ denoting the number of dimensions. For mesons there is one relative
coordinate, hence $\rho \sim m^{2}$, whereas the two relative coordinates in
the baryon give $\rho \sim $ $m^{5}$. Weaker-growing potentials lead to a
faster growth of the number of states, but fall short of the behavior (\ref
{hag}). We know of three approaches yielding behavior (\ref{hag}), both
involving combinatorics of infinitely-many degrees of freedom. These are the
{\em Statistical Bootstrap Model} \cite{hagedorn,SBM1,hag94,SBM2},
Bag Models \cite{MIT:bag,Kapu,Gagnon}, and Dual
String Models \cite{Jacob}. The first two, however, lead to the
same rate of growth for the mesons and baryons. Statistical
Bootstrap Models are discussed in Sec. \ref{SMMsec}.
In Bag Models \cite{MIT:bag,Kapu,Gagnon} the exponential growth
of the spectrum is associated
with the melting out of the vacuum around the bag when the hadron is
being excited. Since the scales in the Bag Model are
practically the same for the meson
and the baryon (the size scales as the number
of constituents to the power
$1/4$), the Bag Models are not capable of answering question 2.
On the other hand, the {\em %
Dual String Models} \cite{Jacob} is offer a natural explanation
of questions 1 and 2. This has already been pointed out in Ref. \cite{tmb}.

\subsection{Statistical Bootstrap Models}

\label{SMMsec} {\em Statistical bootstrap} models \cite{hagedorn,hag94,SBM2}
form particles from clusters of particles, and employ the principle of
self-similarity. The simplest, ``generic'', bootstrap equation has the form
\begin{equation}
\rho (m)=\delta (m-m_{0})+\sum_{n=2}^{\infty }\frac{1}{n!}\int_{0}^{\infty
}dm_{1}...dm_{n}\times \delta (m-\sum_{i=1}^{n}m_{i})\rho (m_{1})...\rho
(m_{n}),  \label{boot}
\end{equation}
where $\rho (m)$ is the particle spectrum (here, for a moment, mesons and
baryons are not distinguished). Equation (\ref{boot}) can be nicely solved
with help of Laplace transforms \cite{hagedorn,hag94,yellin}, yielding the
asymptotic solution $\rho (m)\sim \exp (m/T)$, with $T={m_{0}}/{\log(-\log
\frac{4}{e})}$. More complicated bootstrap equations involve integration
over momenta, more degrees of freedom, different combinatorial factors \cite
{hag94}, however, irrespectively of these details, they always lead to an
exponentially growing spectrum. It can be shown, following {\em e.g.} the
steps of Ref. \cite{Nahm}, that the model leads to equal Hagedorn
temperatures for mesons and for baryons. This is quite obvious. Since
baryons are formed by attaching mesons to the ``input'' baryon, the baryon
spectrum grows at exactly the same rate as the meson spectrum. Specific
calculations confirm this simple observation. Thus the bootstrap idea {\em %
is not capable} of explaining the different behavior of mesons and baryons
in Fig. \ref{cumul}.

\subsection{Dual String models}

\label{dual}

\begin{figure}[tbp]
\centerline{%
\psfig{figure=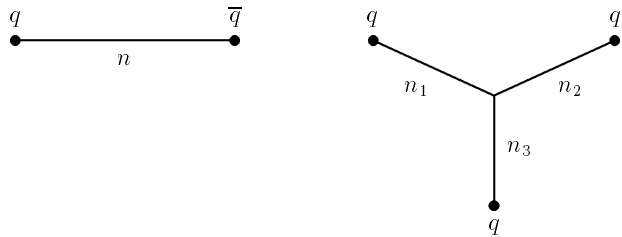,height=4.0cm,bbllx=149bp,bblly=529bp,bburx=457bp,bbury=665bp,clip=}}
\vspace{-2mm}
\caption{Meson and baryon string configurations.}
\label{strcon}
\end{figure}

The {\em Dual String models} \cite{Jacob} also date back to pre-QCD times.
Their greatest success is a natural explanation of the Regge trajectories --
a basic experimental fact which remain a serious problem for other
approaches. Similarly to the bootstrap models, the Dual String Models lead
to exponentially-growing spectra, but they do give the demanded effect of $%
T_{{\rm meson}}>T_{{\rm baryon}}$, at least at asymptotic masses \cite{tmb}.

Let us analyze mesons first. The particle spectrum is generated by the
harmonic-oscillator operator describing vibrations of the string, 
\begin{equation}
N=\sum_{k=1}^{\infty }\sum_{\mu =1}^{D}ka_{k,\mu }^{\dagger }a_{k,\mu },
\label{ho}
\end{equation}
where $k$ labels the modes and $\mu $ labels additional degeneracy, related
to the number of dimensions \cite{Jacob}. Eigenvalues of $N$ are composed in
order to get the square of mass of the meson, according to the Regge formula 
\begin{equation}
\alpha ^{\prime }m^{2}-\alpha _{0}=n,  \label{regge}
\end{equation}
where $\alpha ^{\prime }\sim 1{\rm GeV}^{-2}$ is the Regge slope, and $%
\alpha _{0}\approx 0$ is the intercept. Here is an example: take $n=5$. The
value 5 can be formed by taking the $k=5$ eigenvalue of $N$ (this is the
leading Regge trajectory, with a maximum angular momentum), but we can also
obtain the same $m^{2}$ by exciting one $k=4$ and one $k=1$ mode,
alternatively $k=3$ and $k=2$ modes, and so on. The number of possibilities
corresponds to partitioning the number $5$ into natural components: 5, 4+1,
3+2, 3+1+1, 2+2+1, 2+1+1+1, 1+1+1+1+1. Here we have 7 possibilities, but the
number of partitions grows very fast with $n$. Partitions with more than one
component describe the sub-leading Regge trajectories. With $D$ degrees of
freedom each component can come in $D$ different species. Let us denote the
number of partitions in our problem as $P_{D}(n)$. For large $n$ the
asymptotic formula for {\em partitio numerorum} leads to the exponential
spectrum according to the formula \cite{partitio,Jacob}. 
\begin{equation}
\rho(m)=2\alpha ^{\prime }mP_{D}(n),\quad P_{D}(n)\simeq \sqrt{\frac{1}{2n}}%
\left( \frac{D}{24n}\right) ^{\frac{D+1}{4}}\exp \left( 2\pi \sqrt{\frac{Dn}{%
6}}\right) ,  \label{parti}
\end{equation}
where $n=\alpha ^{\prime }m^{2}$. We can now read-off the mesonic Hagedorn
temperature: 
\begin{equation}
T_{{\rm meson}}=\frac{1}{2\pi }\sqrt{\frac{6}{D\alpha^{\prime}}}.
\label{tmes}
\end{equation}

\begin{figure}[tbp]
\centerline{%
\psfig{figure=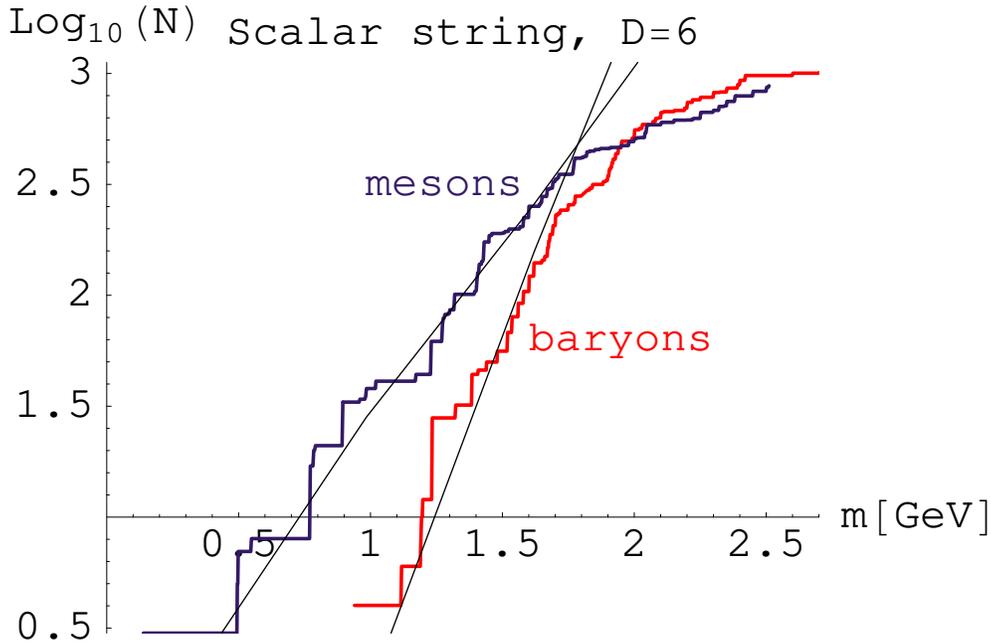,height=9cm,bbllx=23bp,bblly=450bp,bburx=480bp,bbury=758bp,clip=}}
\vspace{0mm}
\caption{Predicions of the scalar string model of 
Ref. \protect{\cite{dienes}}, with $D=6$.}
\label{ss}
\end{figure}

Now the baryons: the ``Mercedes-Benz'' string configuration for the baryon
is shown in Fig. \ref{strcon}. The three strings vibrate {\em independently}%
, and the corresponding vibration operators, $N$, add up. Consequently,
their eigenvalues $n_{1}$, $n_{2}$, and $n_{3}$ add up. Thus we simply have
a partition problem with $3$ times more degrees of freedom than in the
meson. The replacement $D\rightarrow 3D$ in (\ref{parti}) leads immediately
to 
\begin{equation}
T_{{\rm baryon}}=\frac{1}{2\pi }\sqrt{\frac{2}{D\alpha ^{\prime }}},
\label{tbar}
\end{equation}
such that 
\begin{equation}
T_{{\rm meson}}/T_{{\rm baryon}}=\sqrt{3}.  \label{ratio}
\end{equation}
We stress that the presented picture is fully consistent with the Regge
phenomenology. The leading Regge trajectory for baryons is generated by the
excitation of a single string, {\em i.e. } two out of three numbers $n_{i}$
vanish (this is the quark-diquark configuration). The subleading
trajectories for baryons come in a much larger degeneracy than for mesons,
due to more combinatorial possibilities. The slopes of the meson and baryon
trajectories are universal, and given by $\alpha ^{\prime }$. We stress that
the ``number-of-strings'' mechanism described above is asymptotic. Thus,
there is a problem in applying string models to the experimentally
accessible range of $m$. This range is not asymptotic enough to use Eq. (\ref
{parti}). From the Regge formula (\ref{regge}) we find immediately that for $%
m$ in the range $1-2$GeV the values of $n$ lie between $1$ and $4$, hence $n$
is not large enough to justify the form (\ref{parti}).

One can do better by using an improved asymptotic formula, derived in Ref. 
\cite{dienes}. The results obtained in the scalar string model \cite{dienes}
are displayed in Fig. \ref{ss}. Here the formula for the meson spectrum is 
\begin{equation}
\rho _{{\rm mes}}(m)=36\times \rho _{{\em scalar}}(m),\;\;\;\rho _{{\em %
scalar}}(m)=\frac{2\alpha ^{\prime }}{(4\pi \alpha ^{\prime }m T_{{\rm mes}%
})^{\nu }}{mI_{\nu}(\frac{m}{T_{{\rm mes}}})},  \label{scals}
\end{equation}
where $I_{2}$ is a modified Bessel function, $T_{{\rm mes}}$ is the meson
Hagedorn temperature (the {\em only} adjustable parameter here), and $\nu
=1+D/2$, with $D$ denoting the number of transverse dimensions. The factor
of $36=6\times 6$ is just the spin-flavor degeneracy of the $\overline{q}q$
configuration \cite{dienes}. For the baryons we fold the three scalar-string
densities, $\rho _{{\em scalar}}(m)$. We use $56$ (rather than $36$) copies
of the string, which is the degeneracy of the baryon multiplet in the ground
state. We notice good agreement with data in Fig. \ref{ss}, for $D=6$. Note
that both curves are fitted with only one parameter, $T_{{\rm mes}}$. For
lower values of $D$ one can fit the mesons equally well, but too many baryon
states are predicted.

\subsection{Exotics as dual strings}

\begin{figure}[tbp]
\centerline{%
\psfig{figure=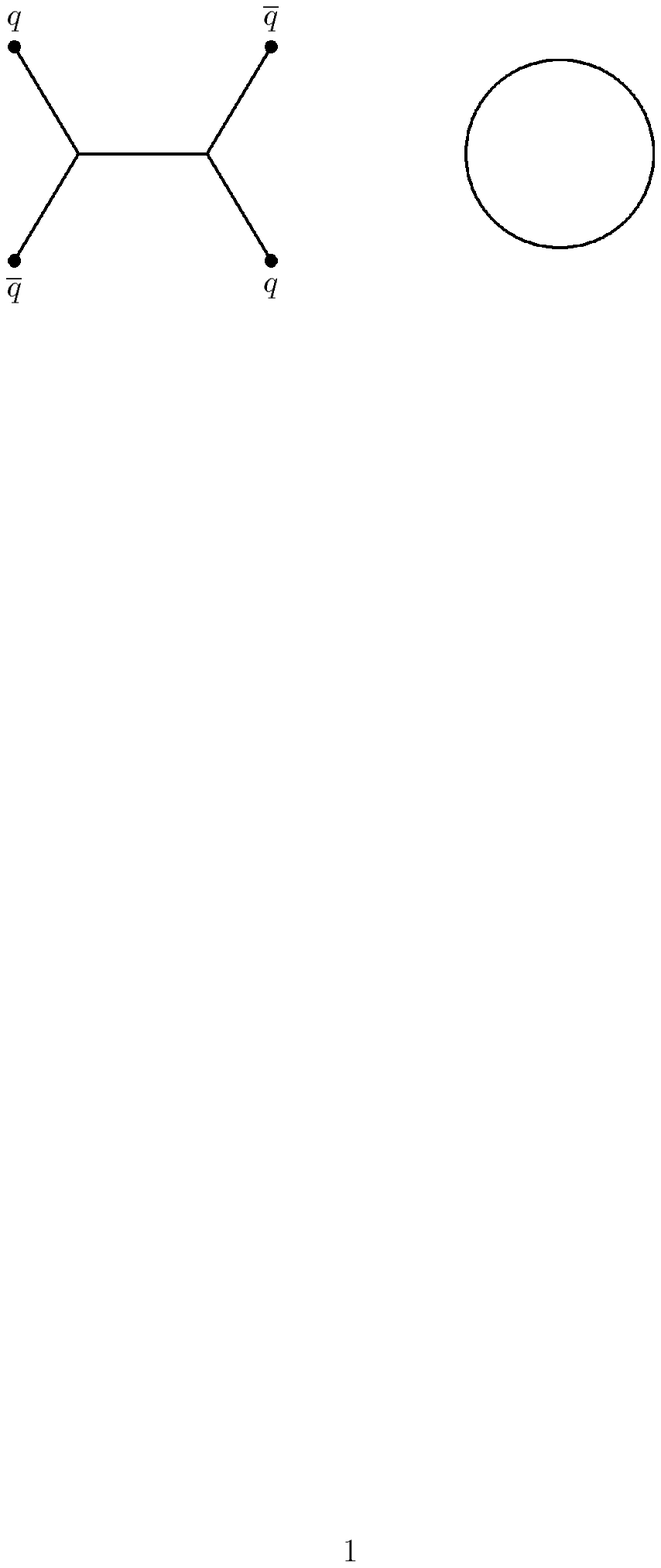,height=4.0cm,bbllx=150bp,bblly=600bp,bburx=435bp,bbury=735bp,clip=}}
\vspace{-2mm}
\caption{$\overline{q}q\overline{q}q$ and glueball configurations.}
\label{exo}
\end{figure}

During this workshop we have heard many talks on hadron exotics. If an
exotic is a multi-string configuration, {\em e.g.} as in Fig. \ref{exo},
then the corresponding spectrum will grow exponentially with the Hagedorn
temperature inversely proportional to the square root of the number of
strings. For instance, $T_{q\overline{q}q\overline{q}}=\frac{1}{\sqrt{5}}T_{%
{\rm meson}}$. This is reminiscent of the effect described in Ref. \cite
{cudell}. For the glueballs, described by the closed string in Fig. (\ref
{exo}), we get $T_G=T_{{\rm meson}}$.

Thus, according to the string model, the $q\overline{q}q\overline{q}$ grow
more rapidly than non-exotic mesons and baryons, and glueballs grow at the
same rate as mesons.

\section{Other approaches}

\label{otherapp}

In the remaining part of this talk we will, in a sense, work against our
results presented in previous sections, where have we argued that the plots
of Fig. \ref{cumul} are linear, and offered an explanation of the difference
between the mesonic and baryonic Hagedorn temperatures within the Dual
String Models.

What if the experimental plots of Fig. \ref{cumul} are not really linear,
and the effect of bending down of the curves at higher masses is physical,
rather than due to incomplete experiments? Below we will show alternative
descriptions which do not comply to Eq. (\ref{hag}), but nevertheless
reproduce the present data at least as good as the Hagedorn-like fits.

\subsection{Compound hadrons}

In the statistical model of nuclear reactions one uses the {\em %
compound-nucleus model} \cite{compound,BMot1}. In this model the density of
states grows at large excitation energies, $E^*$, according to the formula 
\begin{equation}
\rho(E)\sim (E^*)^{-5/4} e^{a \sqrt{E^*}},  \label{comn}
\end{equation}
where $a$ is a constant. Formula (\ref{comn}) can be derived within the
Fermi gas model \cite{BMot1}. More generally, it can be derived in a model
where the single-particle orbits are {\em equally spaced}. One then
considers $1p1h$, $2p2h$, $3p3h$, {\em etc.}, excitations and counts the
number of states at a given excitation energy, $E^*$. Amusingly, this leads 
\cite{zen} to the partitio numerorum formula (\ref{parti}), but now the
number $n$ has the interpretation $n=E^*/\Delta E$, with $\Delta E$ denoting
the level spacing.

\begin{figure}[tbp]
\centerline{%
\psfig{figure=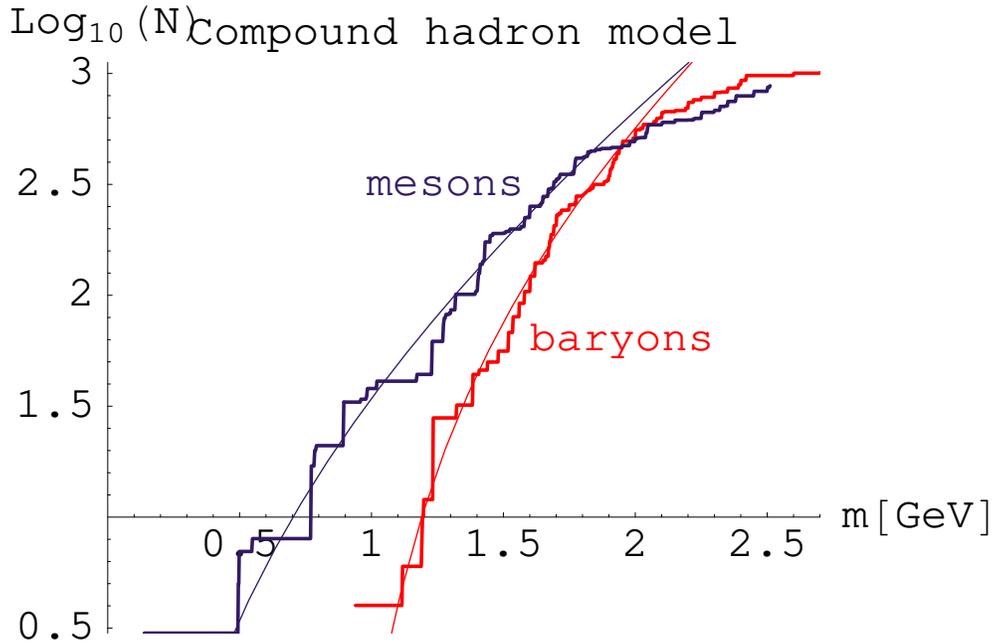,height=9cm,bbllx=23bp,bblly=450bp,bburx=480bp,bbury=758bp,clip=}}
\vspace{0mm}
\caption{Fits of the Compound Hadron Model, Eq. (\protect{\ref{chmrho}}).}
\label{chm}
\end{figure}

We now use the following Compound-Hadron-Model formula for the mass spectra: 
\begin{equation}
\rho (m)=\frac{A\Theta (m-m_{0})\exp \left( 2\pi \sqrt{\frac{\left(
m-m_{0}\right) }{6\Delta E}}\right) }{\left( \left( m-m_{0}\right) ^{2}+(0.5%
{\rm GeV})^{2}\right) ^{5/8}},  \label{chmrho}
\end{equation}
where $A$ is a constant, $m_{0}$ is the ground-state mass, and $\Delta E$ is
the average level spacing. The constant $0.5$GeV in the denominator has been
introduced {\em ad hoc}, similarly as in Eq. (\ref{HR}), in order for the
formula to make sense at $m\rightarrow m_{0}$. Asymptotically, the power of $%
m$ multiplying the exponent is $-5/4$, as in Eq. (\ref{comn}).

The underlying physical picture behind compound hadrons is as follows:
hadrons are bound objects of constituents (quarks, gluons, pions). The Fock
space contains a ground state, and excitations on top of it. In the case of
the compound nucleus these elementary excitations are $1p1h$, $2p2h$, $3p3h,$
{\em etc.} states. In the case of hadrons they are formed of $q\bar{q}$ and
gluon excitations, {\em e.g.} for mesons we have $q\bar{q}$, $q\bar{q}g$, $qq%
\bar{q}\bar{q}$, $q\bar{q}gg$, {\em etc.} We can form the excitation energy
(hadron mass) by differently composing elementary excitations. This bring us
to the above-described combinatorial problem \cite{zen}. It seems reasonable
to take zero ground-state energy for mesons, $m_{0}^{{\rm mes}}=0$, since
they are excitations on top of the vacuum. For baryons we take $m_{0}^{{\rm %
bar}}=900$MeV, which is the mass of the nucleon. The quantity $\Delta E$ is
treated as a model parameter and is fitted to data.

The results of the compound-hadron-model fit, Eq. (\ref{chmrho}), are shown
in Fig. \ref{chm}. The curves are slightly bent down, compared to the
Hagedorn-like fits of Figs. \ref{cumul},\ref{varhag}, which is caused by the
square root in the exponent of Eq. (\ref{chmrho}). But the fits are at least
as good, or even better when the fit region is extended to $m=2$GeV.
Numerically, the least-square fit for $m$ up to 1.8GeV gives $\Delta E^{{\rm %
mes}}=100$MeV for mesons, and $\Delta E^{{\rm bar}}=106$MeV for baryons. The
proximity of these numbers shows that the scales for mesons and baryons are
similar, as should be the case.

The obtained values for $\Delta E^{{\rm mes}}$ mean that the corresponding $%
n $ at $m=1.8$GeV is around $18$ for mesons and $9$ for baryons. Such values
of $n$ are sufficiently large to justify the use of the asymptotic formulas.

\subsection{Combinatorial saturation and the light-flavor-desert hypothesis}

\begin{figure}[tbp]
\centerline{%
\psfig{figure=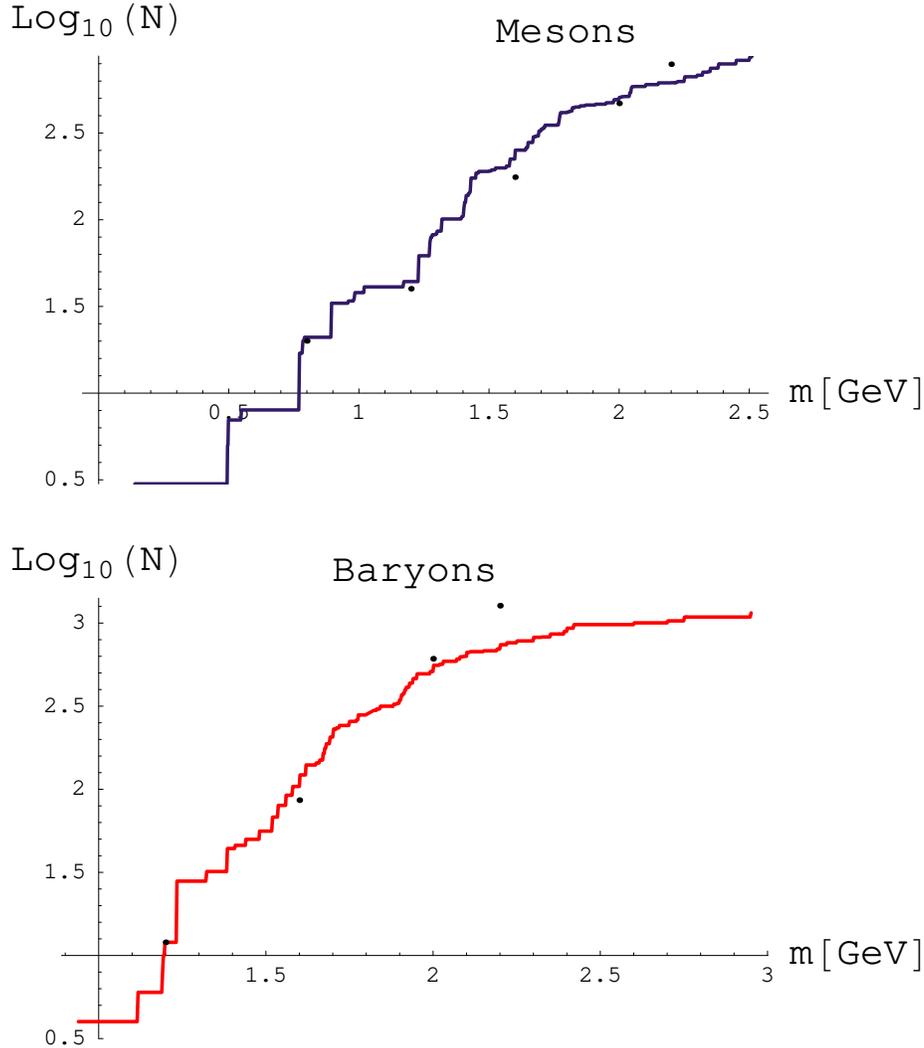,height=14.2cm,bbllx=42bp,bblly=214bp,bburx=520bp,bbury=752bp,clip=}}
\vspace{0mm}
\caption{Experimental cumulants and the predicions of the quark model of
Ref. {\protect\cite{godfrey,capstick}}, as counted in Ref. {\protect\cite
{freund}}, indicated by dots.}
\label{qm}
\end{figure}

There is a possibility of an interesting effect we wish to point out. It is
natural to expect that a bound hadronic system has an upper limit for the
excitation energy. It is helpful to think here of bags of finite depth.
Thus, in constructing the single-particle Fock space for bound objects we
should have a limited number of quanta to our disposal. If such a limit is
put into the Compound Hadron Model, it will result in a maximum number of
states that can possibly be formed out of light quarks \cite{twoT}. We can
call it the {\em ``light-flavor-desert hypothesis''}: above a certain mass
there are no more light-flavor resonances. Certainly, this is tangential to
the conventional wisdom that the Regge trajectories should continue
indefinitely. Note, however, that infinite Regge trajectories have recently
been challenged by Brisudov\'{a}, Burakovsky and Goldman, who claim that
they should stop around $m\sim 2.7$GeV. Amusingly, this is consistent with
the presently-available data. The cumulants if Fig. \ref{cumul} flatten-out
in that region.

\subsection{Quark models}

Many talk in this workshop were devoted to variants of the quark model. Here
we present the result of counting of states in the model of Refs. \cite
{godfrey,capstick}, as made by Freund and Rosner \cite{freund}.

When we look at Fig. \ref{qm}, we again see good agreement in the predicted
and experimental number of states. This is not at all surprising, since the
quark model is designed to fit the data ``state by state'' in the low-mass
regime. As for other approaches, spectra at higher $m$ would be needed to
verify the predictions.

\section{Final remarks}

There are many fundamental questions which should be cleared when more
experimental data on hadron resonances are available: Is the Hagedorn
hypothesis of exponentially-growing spectra indeed correct, or is the growth
weaker at higher masses? Do the Regge trajectories continue for ever, or
stop? Consequently, is there a light-flavor desert above a certain mass? Are
there exotic states, if so, at what rate do they grow?... Certainly, the
spectrum above 2GeV may reveal many answers and help us to verify various
models and approaches.

However, even the presently-available spectrum allows for interesting
speculations. Recall the remarks made here by Leonid Glozman, concerning the
parity doublets in the $N$ and $\Delta$ spectra above $2$GeV \cite{glozman}.
Almost all states in that region can be paired, and such a regularity
suggests that the data in that region may be complete! This, in turn,
indicates that the bending down of the cumulants in Fig. \ref{cumul} may be
a physical, rather than experimental effect.

Another important aspect, not touched in this talk, are the thermodynamical
implications of the presence of two distinct Hagedorn temperatures for the
phenomenology of heavy-ion collisions, transition to quark-gluon plasma, 
{\em etc.} This will be discussed in \cite{zen}.

The author thanks Keith R. Dienes for many profitable e-mail discussions on
the issues of hadron spectra in string models, as well as to Andrzej Bia\l{}%
as, Andrzej Horzela, Jan Kwieci\'nski, and Kacper Zalewski for numerous
useful comments and encouragement.


\begin{thebibliography}{10}

\bibitem{hagedorn}
R. Hagedorn, Suppl. Nuovo Cim. 3 (1965) 147

\bibitem{SBM1}
R. Hagedorn, CERN preprint No.~{CERN 71-12} (1971), and references therein

\bibitem{hag94}
R. Hagedorn, CERN preprint No.~{CERN-TH.7190/94} (1994), and references therein

\bibitem{PDG}
{Particle Data Group}, Eur. Phys. J. C 3 (1998) 1

\bibitem{twoT}
W. Broniowski and W. Florkowski, INP Cracow preprint No.~1843/PH (2000),
  {\tt hep-ph//0004104}

\bibitem{tmb}
W. Broniowski, talk presented at {\em Meson 2000}, 19-23
  May 2000, Cracow, {\tt hep-ph//0006020}

\bibitem{Jacob}
 in {Dual Theory}, Vol.~1 of {Physics Reports reprint book series}, edited by
  M. Jacob (North Holland, Amsterdam, 1974)

\bibitem{HagRan}
R. Hagedorn and J. Ranft, Suppl. Nuovo Cim. 6 (1968) 169

\bibitem{freund}
G.~O. Freund and J.~L. Rosner, {Phys. Rev. Lett.} 68 (1992) 765

\bibitem{dienes}
K.~R. Dienes and J.-R. Cudell, {Phys. Rev. Lett.} 72 (1994) 187

\bibitem{SBM2}
S. Frautschi, {Phys. Rev. {\bf D}} 3 (1971) 2821

\bibitem{MIT:bag}
A. Chodos, R. L. Jaffe, K. Johnson, C. B. Thorn, and V. F. Weisskopf,
{Phys. Rev. {\bf D}} 9 (1974) 3471

\bibitem{Kapu}
J. Kapusta, {Nucl. Phys. {\bf B}} 196 (1982) 1

\bibitem{Gagnon}
R. Gagnon and L. Marleau, {Phys. Rev. {\bf D}} 35 (1987) 910

\bibitem{yellin}
J. Yellin, {Nucl. Phys. {\bf B}} 52 (1973) 583

\bibitem{Nahm}
W. Nahm, {Nucl. Phys. {\bf B}} 45 (1972) 525

\bibitem{partitio}
G.~H. Hardy and S.~S. Ramanujan, Proc. London Math. Soc. 17 (1918) 76

\bibitem{cudell}
J.-R. Cudell and K.~R. Dienes, {Phys. Rev. Lett.} 69 (1992) 1324

\bibitem{compound}
E. Vogt,  in {Advances in Nuclear Physics}, edited by M. Baranger and E. Vogt
  (Plenum Press, New York, 1968), Vol.~1, p.\ 261

\bibitem{BMot1}
A. Bohr and B.~R. Mottelson, Nuclear Structure (Benjamin, Reading,
  Massachussetts, 1969), Vol.~1

\bibitem{zen}
W. Broniowski, W. Florkowski, and P. {\.Z}enczykowski, {to be published}

\bibitem{godfrey}
S. Godfrey and N. Isgur, {Phys. Rev. {\bf D}} 32 (1985) 189

\bibitem{capstick}
S. Capstick and N. Isgur, {Phys. Rev. {\bf D}} 34 (1986) 2809

\bibitem{glozman}
L.~Y. Glozman, {Nucl. Phys. {\bf B}} 475 (2000) 329

\end{thebibliography}

\end{document}